%% file: iopconfser-template.tex
\documentclass{iopconfser}

\usepackage{fontawesome5}
\usepackage{amsmath}
\usepackage{amssymb}
\usepackage{booktabs}
\usepackage{multirow}

\begin{document}

\title{Cross-Modal Alignment between Visual Stimuli and Neural Responses in the Visual Cortex}

\author{Xing Gao$^{1}$, Dazhong Rong$^{2}$ (\faEnvelope{}) and Qinming He$^{2}$}

\affil{$^1$University of Electronic Science and Technology of China, Chengdu, China}
\affil{$^2$Zhejiang University, Hangzhou, China}

\email{rdz98@zju.edu.cn}

\begin{abstract}
Investigating the mapping between visual stimuli and neural responses in the visual cortex contributes to a deeper understanding of biological visual processing mechanisms. Most existing studies characterize this mapping by training models to directly encode visual stimuli into neural responses or decode neural responses into visual stimuli. However, due to neural response variability and limited neural recording techniques, these studies suffer from overfitting and lack generalizability. Motivated by this challenge, in this paper we shift the tasks from conventional direct encoding and decoding to discriminative encoding and decoding, which are more reasonable. And on top of this we propose a cross-modal alignment approach, named Visual-Neural Alignment (VNA). To thoroughly test the performance of the three methods (direct encoding, direct decoding, and our proposed VNA) on discriminative encoding and decoding tasks, we conduct extensive experiments on three invasive visual cortex datasets, involving two types of subject mammals (mice and macaques). The results demonstrate that our VNA generally outperforms direct encoding and direct decoding, indicating our VNA can most precisely characterize the above visual-neural mapping among the three methods.
\end{abstract}

\input{sections/introduction}
\input{sections/background}
\input{sections/methods}
\input{sections/results}

\section{Conclusion and Future Work}
In this work, we proposed a Visual-Neural Alignment framework that jointly models the bidirectional relationship between visual stimuli and brain activity. By leveraging a shared embedding space and contrastive training, our method achieves strong performance on both discriminative encoding and decoding tasks across multiple datasets and species. Compared with direct encoding and decoding baselines, our approach demonstrates superior accuracy and generalization, highlighting the benefits of learning aligned representations. For future work, we plan to extend our framework to support temporally dynamic stimuli and neural responses, enabling applications in video-based neural decoding and brain-state tracking. 

\bibliographystyle{splncs04}
\bibliography{refs}

\end{document}

%% file: sections/introduction.tex
\section{Introduction}
Understanding how external visual stimuli are processed and represented by the brain is a fundamental question in neuroscience~\cite{zhuang2021unsupervised,rong2025improving}, with broad implications for both basic science and clinical applications. In particular, uncovering the mapping between external visual stimuli and neural responses in the visual cortex is not only critical for deciphering the principles of biological visual processing, but also has the potential to inform the design of brain-machine interfaces, neural prosthetics, and biologically inspired artificial vision systems which may be more robust to various attacks~\cite{rong2024clean,zhu2024multiview}.

Traditionally, computational approaches to characterize the above visual-neural mapping fall into two main paradigms: direct encoding (predicting neural responses from visual stimuli) and direct decoding (reconstructing visual stimuli from neural responses)~\cite{nguyen2024estimating,chen2024decoding}. These approaches, often implemented using sophisticated regression or deep learning models~\cite{ma2023using,wei2024speed}, have provided valuable insights into neural representational spaces. While these methods have achieved some success, they suffer from fundamental limitations rooted in biological and technical constraints, making them unreliable or poorly generalizable in practice.

Specifically, direct encoding models are fundamentally challenged by the variability of neural responses. For the same visual stimulus (e.g., a single image), the brain may produce multiple distinct neural response patterns across trials, even under identical experimental conditions. This stochasticity arises from both intrinsic neural noise and unobserved internal states. As a result, encoding models often overfit to one particular observed response pattern, learning spurious associations between an image and a single neural response, rather than discovering the more fundamental, generalizable rules that govern the visual-to-neural transformation.

On the other hand, direct decoding models are severely constrained by the limitations of current neural recording techniques. With invasive electrophysiology, one can typically record responses from only a few hundred or a few thousand neurons simultaneously, whereas millions of neurons across multiple areas of the visual cortex contribute to stimulus processing. This creates a significant information bottleneck: the recorded neural data simply do not contain sufficient information to uniquely reconstruct the original stimulus. In practice, decoding models often rely on memorizing correlations between particular neural patterns and specific visual stimuli, rather than genuinely reconstructing the visual input from a sufficiently informative neural representation.

Motivated by these challenges, we propose a conceptual shift from direct encoding and decoding to more robust and biologically plausible tasks we term discriminative encoding and decoding. Rather than aiming to exactly predict neural responses or reconstruct images, discriminative tasks focus on identifying or distinguishing between conditions (e.g., which of several stimuli was presented, or which neural response corresponds to a given image). This shift alleviates the burden on the model to capture noisy high-dimensional mappings and instead emphasizes information-preserving transformations that are more stable and generalizable. This shift better aligns with the inherently noisy and lossy nature of neural data and avoids over-reliance on unstable one-to-one mappings.

Building on the idea of discriminative encoding and decoding, we introduce a novel cross-modal alignment approach named Visual-Neural Alignment (VNA). This method aligns visual and neural representations in a shared latent space by contrastive learning to enhance discriminative power. To comprehensively evaluate the effectiveness of VNA, we benchmark it against standard direct encoding and decoding models on discriminative encoding and decoding tasks. Our experiments span three invasive visual cortex datasets involving two species (\textit{i.e.,} mice and macaques). The experimental results show that VNA consistently outperforms baseline methods in discriminative encoding and decoding settings, suggesting that cross-modal alignment provides a more precise and robust characterization of the visual-neural relationship.

The main contributions of this paper can be summarized as following:
\begin{enumerate}
    \item We propose a new framework named Visual-Neural Alignment (VNA) to align the features extracted from the two modalities (\textit{i.e.,} external visual stimuli and internal neural responses) into a shared latent space by contrastive learning.
    \item We design two tasks (\textit{i.e.,} discriminative visual encoding and decoding) to quantitatively evaluate how well a model captures the mapping between visual stimuli and neural responses in the visual cortex. Compared with conventional visual encoding and decoding tasks, which aim to directly encode visual stimuli into neural responses or directly decode visual stimuli from neural responses, our designed discriminative tasks are more reasonable due to neural variability and limited information about visual stimuli carried by neural responses.
    \item We conduct extensive experiments to evaluate the performance of our VNA framework on the discriminative visual encoding and decoding tasks. The experimental results demonstrate the superiority of our VNA over direct encoding and decoding. 
\end{enumerate}

%% file: sections/background.tex
\section{Background}
Research on brain visual encoding and decoding has largely benefited from advances in neural signal acquisition techniques. Depending on the level of invasiveness, spatial and temporal resolution, and signal type, the methods used to capture brain activity fall into three main categories: non-invasive electroencephalography (EEG), non-invasive functional magnetic resonance imaging (fMRI), and invasive electrophysiology (e.g., multi-electrode arrays, Neuropixels probes). Each of these modalities has enabled progress in understanding visual processing, yet they present distinct modeling challenges and are suitable for different types of computational approaches.

\subsection{EEG-based Visual Encoding and Decoding}
EEG measures electrical potentials at the scalp surface with millisecond-level temporal resolution but relatively poor spatial localization. Its non-invasive nature makes it widely accessible for brain-computer interface (BCI) research and real-time neural decoding systems. In the context of visual stimuli, EEG has been used to classify perceived object categories, reconstruct low-resolution images, and decode attention-related features. 

Typical methods in EEG decoding include regularized linear models, convolutional neural networks, and recently, transformer-based approaches for temporal feature extraction. However, the limited spatial specificity of EEG signals constrains its utility in high-resolution visual-neural mapping.

\subsection{fMRI-based Visual Encoding and Decoding}
Functional MRI records the blood-oxygen-level-dependent (BOLD) signals that indirectly reflect neural activity. Compared to EEG, fMRI offers much higher spatial resolution, capturing voxel-wise activity across the whole brain, albeit with much slower temporal dynamics (on the order of seconds). This makes fMRI particularly suitable for modeling static visual representations in regions like V1--V4 and inferotemporal cortex.

Encoding models using fMRI data have been highly influential. Early works employed Gabor-based models, while more recent approaches utilize deep convolutional neural networks (CNNs) as priors or feature extractors for predicting voxel responses. On the decoding side, several studies have demonstrated image reconstruction from fMRI, including natural images, faces, and even imagination. However, due to the sluggish and indirect nature of BOLD signals, fMRI decoding is limited to relatively static or low-frame-rate stimuli, and requires heavy spatial averaging to achieve stable results. 

\subsection{Invasive Electrophysiology-based Visual Encoding and Decoding}
Invasive recordings (including single-unit recordings, multi-electrode arrays, and recently Neuropixels probes) offer millisecond-level temporal resolution and cellular-level spatial resolution. These methods directly capture spiking activity and local field potentials from neurons in the visual cortex, making them ideal for fine-grained, trial-level modeling of visual processing. While limited to animal models (e.g., rodents and non-human primates), invasive datasets provide the richest and most detailed access to the visual information encoded in the brain.

Several recent works have developed encoding models based on deep neural networks trained to predict spiking responses from natural images, revealing tuning properties of V1 and higher areas. On the decoding side, however, accurate stimulus reconstruction remains extremely challenging, largely due to the small number of recorded neurons compared to the full neural population involved in visual processing. Some efforts have explored population vector approaches or probabilistic decoding, but most decoding models struggle to generalize across images or sessions, especially when trained directly to reconstruct input stimuli.

Our work falls into this invasive domain, but differs significantly in its focus on discriminative encoding and decoding tasks, rather than full regression or reconstruction. Furthermore, we propose a cross-modal alignment framework to better capture shared structure between visual and neural domains, rather than learning one-to-one deterministic mappings that are often fragile in the presence of neural variability or recording constraints.

%% file: sections/methods.tex
\section{Methods} % 2页+
The framework of our work involves two sub-models, a visual encoder and a neural encoder, extracting features from visual stimuli and neural spikes respectively.
Inspired by recent success on learning transferable visual models from natural language supervision~\cite{radford2021learning}, we similarly adopt contrastive learning to align the features of the two different modalities (\textit{i.e.,} visual stimuli and neural spikes) into the shared latent space.
After the contrastive learning, the full model can effectively perform discriminative visual encoding and decoding tasks.
In the following, we first propose our overall framework, and then introduce our designed discriminative tasks of visual encoding and decoding.

\subsection{Overall Framework}
\subsubsection{Visual Encoder.}
For each original image of our visual stimuli, we first resize it to $c\times 64\times 64$, where $c$ is the number of channels and $64$ is the number of pixels along each side of the image. The resized image is then passed through a convolutional encoder to extract visual features. Formally, we define the space of input images as $\mathbb{R}^{c\times 64\times 64}$ and the space of extracted latent features as $\mathbb{R}^{d}$, where $d$ is the latent dimensionality. Our visual encoder can be formally defined as a function $f:\mathbb{R}^{c\times 64\times 64}\rightarrow\mathbb{R}^{d}$. Specifically, our visual encoder consists of a stack of five convolutional layers with kernel size 4, stride 2, and padding 1, producing output channels of 16, 32, 64, 128, and 256, respectively. Each convolutional layer is followed by a 2D batch normalization layer and a LeakyReLU activation. After the final convolutional layer, the resulting feature map has spatial size $2 \times 2$ and channel size 256. This $256\times2\times2$ tensor is flattened into a 1024-dimensional vector, which is then projected into the latent space $\mathbb{R}^{d}$ using a linear layer.

\subsubsection{Spike Encoder.}
For each visual stimulus, its corresponding neural response is represented as a firing rate vector, where each element indicates the firing rate of a recorded neuron. We denote the space of raw neural responses as $\mathbb{R}^{n}$, where $n$ is the number of recorded neurons. To extract a compact neural representation, we first pass the $n$-dimensional spike vector through a fully connected layer with output dimension 512, followed by batch normalization and a LeakyReLU activation function. This transformation serves to nonlinearly project the raw neural responses into a higher-level representation space. Subsequently, the resulting 512-dimensional vector is projected into a shared latent space of dimension $d$ using another linear layer. The spike encoder can be defined as a function $g:\mathbb{R}^{n} \rightarrow \mathbb{R}^{d}$.

\subsubsection{Contrastive Learning.} % 突出
% 训练分若干个batch进行，每一个batch中会随机选出N个图像刺激和对应的神经信号。对于每个batch内而言，我们用M_i表示第i张图像，v_i表示第i个神经信号。然后：
The training data is divided into multiple batches, with each batch containing $N$ randomly selected pairs of input images and corresponding neural responses, denoted by $\{\mathbf{M}_i,\mathbf{v}_i\}_{i=1}^N$, where $N$ is the batch size, $\mathbf{M}_i$ and $\mathbf{v}_i$ are the $i$-th input image and the $i$-th neural response in the batch respectively.
% 我们定义X_i和Y_j的相似度分数是s_{ij} = f(X_i), g(Y_i)
We define the similarity score between $\mathbf{M}_i$ and $\mathbf{v}_j$ as the cosine similarity of their features extracted by our image encoder and our spike encoder respectively, denoted by $w_{ij}=\frac{f(\mathbf{M}_i)\cdot g(\mathbf{v}_j)}{\|f(\mathbf{M}_i)\|\cdot\|g(\mathbf{v}_j)\|}$.
% 进一步，我们有$\mathcal{L}=\sum_{i=1}^N s_i$
Based on our defined similarity score, we further define our loss function as:
\begin{equation}
    \mathcal{L}=-\frac{1}{2N}\sum_{i=1}^N\big(\log\frac{\exp(w_{ii})}{\sum_{j=1}^N\exp(w_{ij})}+\log\frac{\exp(w_{ii})}{\sum_{j=1}^N\exp(w_{ji})}\big),
\end{equation}
which aims at boosting the similarity scores between matched images and neural responses, and decreasing the similarity scores between mismatched images and neural responses.
% 在每个batch内，我们都根据上面这个损失函数来优化f和g中可训练的参数。
According to our loss function, we optimize our image encoder and our spike encoder batch by batch.
% 在训练完毕之后，对于一段神经信号v，以及若干个候选图像M_1, M_2, ... M_K,我们怎么根据v挑出最合适的M
After training, given a neural signal $v$ and a set of candidate visual stimuli $\{\mathbf{M}_1, \mathbf{M}_2, \dots,  \mathbf{M}_K \}$, we can predict the corresponding visual stimulus by selecting the image with the highest similarity score to $v$ among the candidates.

\subsection{Discriminative Visual Encoding and Decoding}
To evaluate the alignment between visual stimuli and neural responses, we propose two discriminative tasks: \textit{Discriminative Visual Encoding} and \textit{Discriminative Visual Decoding}. These tasks are designed to assess how well the learned visual and neural representations are aligned in a shared latent space. Unlike generative or regression-based evaluations, our formulation adopts a retrieval-based paradigm, where the model is required to rank candidates and identify the true match.

\subsubsection{Discriminative Visual Encoding.}
In the Discriminative Visual Encoding task, the model is given a visual stimulus image and a set of $K$ candidate neural response vectors. Among these candidates, only one neural response corresponds to the true biological recording evoked by the visual stimulus, while the remaining $K-1$ responses are sampled as distractors. The model computes similarity scores between the image embedding and each of the $K$ neural embeddings, and ranks the candidate responses accordingly.

\subsubsection{Discriminative Visual Decoding.}
In the Discriminative Visual Decoding task, the model is given a neural response vector and $K$ candidate visual images. Only one of these images corresponds to the stimulus that elicited the neural response, and the rest are distractors. The model must compute similarity scores between the neural embedding and each of the $K$ image embeddings, and rank the candidates.

%% file: sections/results.tex
\section{Results}

\subsection{Datasets}
For validating the effectiveness of our proposed method, we adopt three datasets which were collected from mice and macaques under three different types of stimuli. We preprocessed the three datasets following the same procedure as in~\cite{huang2023deep}. In the following, we introduce the three datasets respectively.

\subsubsection{Mouse-movie dataset.}
This dataset~\cite{siegle2021survey} is derived from the Allen Brain Observatory Visual Coding project, in which Neuropixel probes were used to record extracellular neural signals from six distinct visual cortical areas in mice. During data acquisition, mice were repeatedly shown a set of 118 natural scene images, each displayed for 250 ms, in randomized order over 50 trials. For this dataset, $c=1$, $n=800$.

\subsubsection{Macaque-face dataset.}
This dataset~\cite{chang2021explaining} is about the macaque neural spikes of the neurons in the anterior medial (AM) face patch under $2100$ real facial stimuli. We ultimately retain $148$ neurons with a noise ceiling greater than or equal to the threshold of $0.1$. For this dataset, $c=3$, $n=148$.

\subsubsection{Macaque-synthetic dataset.}
This dataset~\cite{majaj2015simple} is composed of macaque $88$ V4 neurons and $168$ IT neurons under $3200$ synthetic visual stimuli, which are generated by mapping the 2D projection of a 3D object model to a natural background. For this dataset, $c=3$, $n=256$.

\subsection{Baselines}
To evaluate the effectiveness of our proposed approach, we compare it against two representative baseline methods: \textit{Direct Encoding} and \textit{Direct Decoding}. Both methods aim to learn a direct mapping between visual stimuli and neural responses, but differ in their directionality and objective.

\subsubsection{Direct Encoding.}
Direct Encoding trains a mapping function $f$ that predicts neural responses from visual stimuli. Formally, given an image $\mathbf{M}$ and its corresponding recorded neural response $\mathbf{v}$, the model $f$ is optimized to minimize the mean squared error (MSE) between the predicted and actual neural signals: $\mathcal{L}_{\text{encode}}=\|f(\mathbf{M})-\mathbf{v}\|_2^2$. In the discriminative encoding task, given a stimulus $\mathbf{M}$ and a set of candidate neural responses $\mathbf{v}_j$, the response whose $\|f(\mathbf{M}) - \mathbf{v}_j\|_2$ is smallest is ranked highest. In the discriminative decoding task, given a recorded response $\mathbf{v}$ and a set of candidate images $\mathbf{M}_i$, each image is scored by the distance $\|f(\mathbf{M}_i)-\mathbf{v}\|_2$, and the image with the smallest discrepancy is ranked highest.

\subsubsection{Direct Decoding.}
Direct Decoding trains a model $g$ that maps neural responses back to visual stimuli. Given a neural response $\mathbf{v}$ and its associated image $\mathbf{M}$, the model $g$ is trained to minimize the MSE between the predicted and ground truth images: $\mathcal{L}_{\text{decode}}=\|\mathbf{M}-g(\mathbf{v})\|_2^2$. This baseline attempts to capture the inverse mapping from neural signals to stimulus space. In the Discriminative Encoding task, given an image $\mathbf{M}$ and a set of candidate responses $\mathbf{v}_j$, each response is evaluated based on $\|g(\mathbf{v}_j) - \mathbf{M}\|_2$, and those closer to the target image are ranked higher. In the Discriminative Decoding task, given a neural response $\mathbf{v}$ and a set of candidate images $\mathbf{M}_i$, the image whose distance to the predicted stimulus $g(\mathbf{v})$ is the smallest is selected as the most likely.

\subsection{Experimental Settings}
In our experiments, we set $N=256$, $K=400$ and $d=64$. We adopt Adam optimizer with learning rate $0.01$ to train our models, and the total number of training epochs is set to $100$ for ensuring the convergence. To quantify model performance on discriminative tasks, we use the \textbf{Area Under the Curve (AUC)} of the receiver operating characteristic (ROC) for binary discrimination between the true match and distractors. For each test instance, the true match is labeled as the positive sample and the remaining $K-1$ distractors are labeled as negatives. AUC measures the probability that the model assigns a higher similarity score to the true match than to a randomly chosen distractor.

\begin{table}[t]
\centering
\caption{\label{tab:exp} Discriminative Task Performance Comparison}
{\begin{tabular}{ccccc}
\toprule
\multirow{2}{*}{Datasets} & \multirow{2}{*}{Methods} & \multicolumn{3}{c}{AUC Metrics} \\
\cmidrule(lr){3-5} ~ & ~ & Encoding & Decoding & Average \\
\midrule
\multirow{3}{*}{Mouse-movie} & Visual-Neural Alignment & \textbf{0.9969} & 0.9973 & \textbf{0.9971} \\
~ & Direct Encoding & 0.9905 & \textbf{0.9982} & 0.9941 \\
~ & Direct Decoding & 0.9956 & 0.9973 & 0.9962 \\
\midrule
\multirow{3}{*}{Macaque-face} & Visual-Neural Alignment & \textbf{0.9104} & 0.9213 & \textbf{0.9126} \\
~ & Direct Encoding & 0.8489 & \textbf{0.9248} & 0.8797 \\
~ & Direct Decoding & 0.8330 & 0.8941 & 0.8601 \\
\midrule
\multirow{3}{*}{Macaque-synthetic} & Visual-Neural Alignment & \textbf{0.9762} & \textbf{0.9771} & \textbf{0.9765} \\
~ & Direct Encoding & 0.8427 & 0.8803 & 0.8582 \\
~ & Direct Decoding & 0.7819 & 0.8437 & 0.8097 \\
\bottomrule
\end{tabular}}
\end{table}

\subsection{Performance Comparison}

We compare our method, Visual-Neural Alignment, with two baselines across three datasets. As shown in Table~\ref{tab:exp}, our method generally outperforms Direct Encoding and Direct Decoding in terms of AUC on both discriminative encoding and decoding tasks. Notably, on all of the three datasets, our approach yields the best average AUC, demonstrating its robustness and superior generalization across species and stimuli types.